\documentclass{aa}  

\usepackage{graphicx}
\usepackage{txfonts}
\usepackage{lipsum}
\usepackage{subcaption}         % necessary for continued figures, example in section 3
\usepackage{lscape}             % to rotate a single page table, example in appendix.
\usepackage{placeins}           % useful with \FloatBarrier, to keep 
\newcommand{\kms}{km\,s$^{-1}$}

\newcommand{\Msun}{$M_{\odot}$}

\begin{document}

   \title{Large-field CO\,(1--0) observations toward the Galactic historical supernova remnants: shocked molecular clouds toward the Crab Nebula}

   %\subtitle{Subtitle}

%%%%%%%%%%%%%%%%%%%%%%%%%%%%%%%%%%%%%%%%
% Please do not include ORCIDs next to author names.
% Only ORCIDs authenticated by individual authors in EDP Sciences editorial system will be taken into account.
% ORCIDs included here will be removed.
%%%%%%%%%%%%%%%%%%%%%%%%%%%%%%%%%%%%%%%%

   \author{Xuepeng~Chen\inst{1,2}, Dong~Wang\inst{1,2}, Qianru~He\inst{1,2}, Jiancheng~Feng\inst{1,2}, Shiyu~Zhang\inst{1}, Li~Sun\inst{1}, \and Yang~Su\inst{1,2}}

   \institute{Purple Mountain Observatory \& Key Laboratory of Radio Astronomy, Chinese Academy of Sciences, 10 Yuanhua Road, 210023 Nanjing, PR China\\
             \email{xpchen@pmo.ac.cn}
             \and School of Astronomy and Space Science, University of Science and Technology of China, Hefei, Anhui 230026, PR China\\ }

   \date{Received August 28th, 2025}
 
  \abstract
   {The investigation of the interstellar gas surrounding the supernova remnants (SNRs) is not only necessary to improve our knowledge of SNRs, 
   but also to understand the nature of the progenitor systems.}
   {As part of the Milky Way Imaging Scroll Painting (MWISP) CO line survey, the aim is to study the interstellar gas toward the Galactic historical SNRs. 
   In this work, we present the CO observational results of the Crab Nebula.} 
   {Using the 3$\times$3 beam Superconducting Spectroscopic Array Receiver (SSAR) at the Purple Mountain Observatory (PMO) 13.7\,m millimeter telescope, 
   we performed large-field and high-sensitivity CO\,(1--0) molecular line observations toward the Crab Nebula. The $Gaia$ optical data are used to measure 
   the distances of molecular clouds detected in this work, while the HI data from the GALFA-HI and the HI4PI surveys are used to make a comparison between 
   the distributions of the molecular and atomic gas.}
   {The CO observations show molecular clouds  toward the Crab Nebula at a velocity range from about 0 to 16\,\kms. After checking the CO spectra, we find 
   shocked signatures in the clouds extending at a velocity of roughly [5, 11]\,\kms. These shocked molecular clouds, with an angular distance of $\sim$\,0\fdg4-0\fdg5 
   toward the Crab Nebula, are located at the shell of a bubble discovered in the HI images at the same velocity range. The dimension of the bubble is roughly 
   2\fdg3\,$\times$\,2\fdg6 and the expansion velocity is $\sim$\,5\,\kms. The kinetic energy released by the supernova is estimated to be about 3.5\,$\times$\,10$^{51}$ erg.}
   {The kinetic energy referred from the shocked molecular clouds, together with the HI bubble, support the picture that the Crab Nebula belongs to a typical core-collapse 
   supernova remnant. Nevertheless, due to the large uncertainty in the distance measurement, further observations are needed to verify the physical association between 
   the shocked molecular clouds and the Crab Nebula.}


\authorrunning{Chen et al.}
\titlerunning{CO observations toward the Crab Nebula}

   \maketitle

%%%%%%%%%%%%%%%%%%%%%%%%%%%%%%%%%%%%%%%%%%%%%%%%%%%%%%%%%%%%%%
\section{Introduction}
%%%%%%%%%%%%%%%%%%%%%%%%%%%%%%%%%%%%%%%%%%%%%%%%%%%%%%%%%%%%%%

The Crab Nebula, the remnant of the historical supernova of 1054 AD, is one of our prime laboratories to study astrophysics and
high-energy physics in the Universe (see the review by Hester 2008). This most-studied SNR is a synchrotron nebula powered 
by the Crab pulsar, enclosed by a bright expanding shell of thermal gas (see reviews by B{\"u}hler \& Blandford 2014 and Amato 
\& Olmi 2021). The observed dimension of the Crab Nebula is approximately 7$'$\,$\times$\,5$'$, with a position angle of 
$\sim$\,126$^\circ$ (measured east of north; e.g., Ng \& Romani 2004). 

Numerous studies have noticed that the kinetic energy of the observed nebula ($\sim$\,3\,$\times$\,10$^{49}$ ergs) is far less 
than the canonical 10$^{51}$ ergs seen in the normal core-collapse supernovae (see Hester 2008 and references therein). It 
was suggested that the Crab Nebula either resulted from an unusual low-energy explosion (e.g., Frail et al. 1995; Yang \& 
Chevalier 2015), or there is a large freely (but undetected yet) expanding shell around the Crab Nebula, which contained the 
most kinetic energy of the supernova explosion (e.g., Chevalier 1977). Whether or not a large shell surrounding the nebula is 
the key to understand the nature of the Crab Nebula, as well as many branches of astrophysics. Nevertheless, despite long 
searches at various wavelengths (e.g., Murdin \& Clark 1981; Frail et al. 1995; Fesen et al. 1997; Wallace et al. 1999; Sollerman 
et al. 2000; Seward et al. 2006), this putative large shell still remains unseen.

In order to study the interstellar gas surrounding the Galactic historical SNRs, we started a program to perform large-field CO
observations toward the known Galactic historical SNRs (see, e.g., Chen et al. 2017), which is part of the Milky Way Imaging 
Scroll Project (MWISP\footnote{http://www.radioast.nsdc.cn/mwisp.php}), an unbiased CO\,(1--0) multi-lines survey toward the 
northern Galactic plane with the PMO 13.7\,m telescope (Su et al. 2019; Sun et al. 2021). 
In this work, we present MWISP CO\,(1--0) observations toward the Crab Nebula. We note that the Crab Nebula was listed in 
the MWISP CO study toward a large catalog of Galactic SNRs (see Zhou et al. 2023). Nevertheless, the larger-field and, in 
particular, higher-sensitivity CO data in this work help us to derive new results toward the remnant.
In Section 2 we describe the observations and data reduction. Observational results are presented and discussed in Section 3 
and summarized in Section 4.

%%%%%%%%%%%%%%%%%%%%%%%%%%%%%%%%%%%
\section{Observations and data reduction}
%%%%%%%%%%%%%%%%%%%%%%%%%%%%%%%%%%%

The CO\,(1--0) observations toward the Crab Nebula were made from 2016 September to 2020 November with the PMO 13.7\,m 
telescope at Delingha in China. A 4$^\circ$\,$\times$\,4$^\circ$ area, covering 183\fdg25\,$\leq$\,$l$\,$\leq$\,187\fdg25 and
 $-$7\fdg25\,$\leq$\,$b$\,$\leq$\,$-$3\fdg25, was observed around the Crab Nebula.
The nine-beam Superconducting Spectroscopic Array Receiver (SSAR; Shan et al 2012) was working as the front end in the 
sideband separation mode in the observations. Three CO\,(1--0) lines were simultaneously observed, $^{12}$CO at the upper 
sideband (USB) and two other lines, $^{13}$CO and C$^{18}$O, at the lower sideband (LSB). Typical system temperatures 
were around 210\,K for the USB and around 130\,K for the LSB, and the variations among different beams are less than 15\%. 
A Fast Fourier Transform (FFT) spectrometer with a total bandwidth of 1\,GHz and 16,384 channels was used as the back end. The 
corresponding velocity resolutions were $\sim$\,0.16\,km\,s$^{-1}$ for the $^{12}$CO line and $\sim$\,0.17\,km\,s$^{-1}$ for both 
the $^{13}$CO and C$^{18}$O lines. 

The observations were preformed via the on-the-fly (OTF) mode. The total of pointing and tracking errors was about 5$''$, while the 
half-power beam width (HPBW) was $\sim$\,52$''$ in the $^{12}$CO line and $\sim$\,54$''$ in the $^{13}$CO and C$^{18}$O lines. 
The main-beam efficiencies during the observations were $\sim$\,44\% for the USB with the differences among the beams less than 
16\%, and $\sim$\,48\% for the LSB with the differences less than 6\%.

In the MWISP survey (see Su et al. 2019 and Sun et al. 2021), the typical rms noises were set to be $\sim$\,0.5\,K for $^{12}$CO 
(at a velocity resolution of $\sim$\,0.16\,\kms) and $\sim$\,0.3\,K for $^{13}$CO and C$^{18}$O (at a velocity resolution of $\sim$\,0.17\,\kms). 
To obtain much higher sensitivity, we deeply mapped a 
1\fdg5\,$\times$\,1\fdg5 area (183\fdg75\,$\leq$\,$l$\,$\leq$\,185\fdg25, $-$6\fdg75\,$\leq$\,$b$\,$\leq$\,$-$5\fdg25) around the 
Crab Nebula. The resulted rms noises in this deep area were about 0.25\,$\pm$\,0.03\,K for the $^{12}$CO line and about 
0.15\,$\pm$\,0.02\,K for the $^{13}$CO and C$^{18}$O lines. Figure\,A.1 in the Appendix shows the distribution of the $^{12}$CO line 
rms nosies in the observations.

After removing bad channels and abnormal spectra, and correcting the first-order (linear) baseline fitting, the observed CO data were 
re-gridded into standard FITS files with a pixel size of 30$\arcsec$\,$\times$\,30$\arcsec$ (approximately half of the beam size). All 
data were reduced using the GILDAS package\footnote{https://www.iram.fr/IRAMFR/GILDAS/} and self-developed pipelines.

%%%%%%%%%%%%%%%%%%%%%%%%%%%%%%%%%%%
\section{Results and discussion}
%%%%%%%%%%%%%%%%%%%%%%%%%%%%%%%%%%%

\subsection{Large-field CO mapping area}

Appendix Figure~A.2 shows the large-field (4$^\circ$\,$\times$\,4$^\circ$) $^{12}$CO velocity channel maps toward the Crab Nebula.
As seen in the channel maps, large-scale molecular cloud, in the velocity range from roughly --2 to 6\,\kms, is observed toward the north of 
the Crab Nebula\footnote{All the directions mentioned in this work are in the Galactic coordinate system.}. Figure~1 shows the $^{12}$CO 
velocity-integrated intensity image. This large-scale molecular cloud was noticed by previous MWISP CO study toward the Crab Nebula (see 
Figure~19 in Zhou et al. 2023).  Zhou et al. (2023) suggested that the cloud might be associated with the Crab Nebula, though no interaction 
evidence was found between each other. 

It is important to investigate the relationship between the Crab Nebula and the large-scale molecular cloud. Figure~2 shows the 
position-velocity (PV) diagrams from the Crab Nebula across the cloud (see PV routings in Fig.\,1). No expanding signature is found 
in the various PV directions (see Fig.\,2). 
We further measure the distance of the cloud using the similar method as described in Yan et al. (2021), based on the MWISP CO 
and Gaia DR3 data (Gaia et al. 2023). This method uses the principle that molecular clouds usually impose higher optical extinction 
than other phases of the ISM. Aided by Bayesian analyses, we derive the distance by identifying the breakpoint in the stellar extinction 
toward molecular cloud (on-cloud region) and using the extinction of Gaia stars around molecular cloud (off-cloud regions) to confirm 
the breakpoint. The systematic error in the measurement is approximately 5\%. The measured distance of the large-scale cloud is
roughly 1374\,pc (see Figure~B.1 in Appendix).

The early estimated distances to the Crab Nebula range from roughly 1400\,pc to 2700\,pc (see Trimble 1968, 1973), and an unweighted 
average of the various methods yields a distance of about 1900\,pc (see Trimble 1973). In the previous studies toward the Crab Nebula,
a distance of 2\,kpc was generally adopted (see, e.g., Hester 2008). Recently, based on the VLBI observations and Gaia DR3 data, the 
distance to the Crab Nebula is more precisely estimated to be 1.90$^{+0.22}_{-0.18}$\,kpc (Lin et al. 2023), which is farther away than
the distance of the north cloud.
Based on the PV diagrams and distances measured in the CO observations, we then consider no association between the Crab Nebula 
and the north large-scale molecular cloud. The detailed study of the large-scale cloud is beyond the scope of this work and will be 
presented in other works in the near future.

\subsection{Deep CO mapping area}

Appendix Figure~A.3 shows the $^{12}$CO velocity channel maps toward the Crab Nebula in the deep observations (see $\S$\,2). 
In the broad bandwidth covering by the MWISP CO survey (16,384 channels or $\pm$1300\,\kms), the $^{12}$CO emission is only detected 
within a velocity range between 0 and +16\,\kms\ toward the Crab region. This is because the Crab Nebula is in the Galactic anti-center direction 
and the velocities of different arms are coupled together (see, e.g., Dame et al. 2001).  As seen in Figure~A.3, small-scale molecular 
clouds are found `around' the remnant. According to the velocity continuity of the CO emission, as well as the similarity of the cloud morphology, 
the CO emission is divided into five velocity ranges: [0, 3], [3, 5], [5, 11], [11, 13], and [13, 16]\,\kms, respectively. Figures~3a \& 3b show 
the CO intensity images integrated for the individual velocity ranges. 

Generally, for a molecular cloud affected by stellar feedback (e.g., stellar wind and/or supernova shock), its observed CO spectra should 
show broadened linewidths and/or asymmetric (non-Gaussian) line profiles (see, e.g., Jiang et al. 2010; Kilpatrick et al. 2016). As seen in 
Appendix Figure~A.4, the CO spectra for the molecular clouds detected in the velocity ranges of [0, 3], [3, 5], [11, 13] and [13, 16]\,\kms\ show 
typical Gaussian line profiles with narrow FWHM linewidths ($\sim$\,1-2\,\kms). It suggests that these clouds are quiescent. Therefore, we 
consider that these molecular clouds are not associated with the Crab Nebula.

On the other hand, the CO spectra of a few molecular clouds in the velocity range of [5, 11]\,\kms\ show line broadening and asymmetric
(see CO spectra Nos.\,6--9 in Figure~4, and also grid spectra in Appendix C). The measured velocity ranges above the baseline 
($\Delta$$v$) of the clouds extend 5--6\,\kms, while the FWHM linewidth ranges from 2 to 4\,\kms. The observed spectra suggest that these 
clouds are shocked. We note that no $^{13}$CO and C$^{18}$O line emission is detected from these clouds. 

We also try to measure the distances of the clouds at the velocity range of [5, 11]\,\kms, using the same method as mentioned above. 
Nevertheless, due to the small angular sizes of the clouds, the number of `on-cloud' stars taken into account is small. Furthermore,
the estimated column density of the clouds range from $\sim$\,5 to $\sim$\,10\,$\times$\,10$^{21}$\,cm$^{-2}$ (a CO-to-H$_2$ conversion 
factor of $X$ = 1.8\,$\times$\,10$^{20}$\,cm$^{-2}$\,K$^{-1}$\,km$^{-1}$\,s is adopted in the estimation; see Dame et al. 2021), which 
means that the optical extinction caused by the clouds is also relatively small. Therefore, large uncertainty remains in the measurements. 
As seen in Figure~B.2, the distance of these clouds is estimated to be 1796$^{+146}_{-121}$\,pc. This distance is comparable to the recently 
measured distance of the Crab Nebula (1.90$^{+0.22}_{-0.18}$\,kpc; Lin et al., 2023). %

%%%%%%%%%%%%%%%%%%%%%%%%%%%%%%%%%%%
\subsection{Molecular clouds shocked by Crab?}
%%%%%%%%%%%%%%%%%%%%%%%%%%%%%%%%%%%

As introduced in Section\,I, the search for the outer shell/shock from the supernova explosion of SN\,1054 has been decades (see, e.g., 
Seward et al. 2006 and Hester 2008, and references therein). Previous HI observations toward the Crab Nebula suggested that the 
remnant is located within a large, low-density bubble (Romani et al. 1990; Wallace et al. 1994, 1999), roughly 90\,pc in radius (at a 
distance of 2\,kpc) and at a velocity range of about $\sim$\,[--17, --3]\,\kms\ (see, e.g., Romani et al. 1990). This bubble is thought to have 
formed as a result of energy input from stellar winds and/or previous supernovae. It is also suggested that such low-density interstellar 
gas environment cloud be the reason of no detection of radio shell toward the Crab (e.g., Frail et al. 1995; Seward et al. 2006).
Dense molecular knots were observed within the Crab Nebula in the high-angular resolution observations (Loh et al. 2010, 2011; 
Wootten et al. 2022). These dense knots, associated with the filaments in the nebula, trace high-velocity expansion of the nebula caused 
by the supernova explosion and pulsar jet. The distribution of the radial velocities of these knots is very asymmetric, and the estimated 
systemic radial velocity is near 0\,\kms\ (see Loh et al. 2011).

Based on the deep CO observations in this work, molecular clouds with shocked signatures are observed toward the southeast of the 
Crab Nebula, with angular distances ranging from $\sim$\,0\fdg4 to $\sim$\,0\fdg5 (see Fig.\, 3b). Although the observed spectral 
linewidths of these clouds ($\Delta$$v$\,$\sim$\,5--6\,\kms\ and FWHM\,$\sim$\,2--4\,\kms; see Fig.\,4) are smaller than the broadened 
velocity linewidths found in a few SNR-MC interaction cases, e.g., IC\,433 ($>$\,25\,\kms; Dickman et al. 1992), the values are comparable 
with many other observed interaction cases, such as G16.7+0.1 ($\Delta$$v$\,$\sim$\,1.5--4.4\,\kms; Reynoso \& Mangum 2000) and 
3C\,397 ($\Delta$$v$\,$\sim$\,4\,\kms\ and FWHM\,$\sim$\,2.5\,\kms; Jiang et al. 2010). Therefore, these clouds could be affected by the 
shocks from the SN\,1054 supernova, which cause the broadened and asymmetric CO spectra seen in the observations.

If the shocks from the SN\,1054 supernova interacted with these molecular clouds, it means that the velocities of the shocks should reach 
at least 16,000\,\kms, referred from the radius of $\sim$\,0\fdg5 (or $\sim$\,17\,pc at a distance of 1.9\,kpc; see Fig.\,3b). This kind of high 
velocities have been observed in a few young SNRs, such as Cas~A ($\sim$\,15,000\,\kms; see, e.g., Fesen et al. 2016) or G1.9+0.3 
($\sim$\,13,000-18,000\,\kms; see, e.g., Borkowski et al. 2013, 2017). 
In this case, the kinetic energy released from the supernova explosion $E_{\rm 0}$ can be estimated from the shock radius $R_{\rm shock}$
by the Sedov-Taylor evolution mode,
\begin{equation}\label{f.energy}
 {\it R_{\rm shock}} \approx 8 \left ( \frac{\it t}{2750\,{\rm yr}}\right )^{\frac{2}{5}} \left ( \frac{\it E_{\rm 0}}{10^{51}\,{\rm erg}}\right )^{\frac{1}{5}} \left ( \frac{\it n_{\rm 0}}{1\,{\rm cm^{-3}}}\right )^{-\frac{1}{5}} {\rm pc},
\end{equation}
where the shock radius $R_{\rm shock}$ is adopted as 17\,pc and the initial gas density $n_0$ as 0.01\,cm$^{-3}$. From the age of SN\,1054, 
the estimated $E_{\rm 0}$ is $\sim$\,3.5\,$\times$\,10$^{51}$\,erg.

Based on the very high-energy $\gamma$-ray observations, recent studies suggest the contribution from hadronic emission from the environment of the Crab 
Nebula by re-accelerated particles (see, e.g., Peng et al. 2022; Spencer et al. 2025), which is somehow consistent with the shocked molecular clouds found 
in this work.
However, it must be noted that the possibility that the molecular clouds are affected by other sources in the region cannot be excluded. For example, 
source SAO\,77293 (or known as HD\,36879), an O7 type massive star, is located in the region (see Fig.\,3b). Its distance, estimated from the Gaia 
parallax data, is roughly 1.8\,kpc. The shocked molecular clouds could be affected by the stellar wind from this source. Nevertheless, molecular clouds 
detected toward SAO\,77293 show quiescent spectra in the CO observations (see No.\,3--5 spectra in Fig.\,4).
Further observations, such as high transition CO observations (or 1720\,MHz OH maser), are needed to verify the shock effect toward the
molecular clouds in the [5, 11]\,\kms\ velocity range.

%%%%%%%%%%%%%%%%%%%%%%%%%%%%%%%%%%%%%%%%
\subsection{A newly-discovered HI bubble associated with the shocked clouds}
%%%%%%%%%%%%%%%%%%%%%%%%%%%%%%%%%%%%%%%%

We also check the HI data from the GALFA-HI (Peek et al. 2018) and HI4PI (HI4PI collaborations et al. 2016) surveys for the Crab Nebula. Figure~5
shows a large-field HI intensity image observed in the GALFA-HI survey. In the similar velocity range of the shocked CO clouds, the GALFA-HI image
clearly shows a bubble surrounding the Crab Nebula (see also HI intensity channel maps in Appendix Fig. D.1), but the pulsar is far (about 1$^\circ$) from 
the center of the bubble ($l$\,=\,184\fdg95, $b$\,=\,--4\fdg95). The dimension of the bubble is roughly 2\fdg3\,$\times$\,2\fdg6, corresponding to a linear 
size of about 80\,pc\,$\times$\,90\,pc (at an assumed distance of 1.9\,kpc). Interestingly, concentric shells are also seen outside the bubble, and the diameter 
of the largest shell is roughly 5$^\circ$ (or 170\,pc).

In the HI PV diagrams across the bubble (see Fig.\,5), cavity-like structures are detected, which indicates that the bubble is expanding. Furthermore, 
spur-like protrusions, spatially coincident with the concentric shells seen in the intensity images, are also observed along the PV diagrams (see Fig.\,5). 
The fitted expansion velocities for the bubble and shells are similar with each other ($\sim$\,5\,$\pm$\,1\,\kms), and the systemic velocity of the bubble 
and shells is roughly 7.5\,\kms. We also note that (1) relatively low angular resolution HI4PI image also show the bubble and the large shell in the same 
velocity range (images not shown here); and (2) no clear bubble or cavity structures is found in the GALFA-HI images (and HI4PI images) in the velocity 
range of [-20, 0]\,\kms\ (see Appendix Fig. D.2).

Figure~6 shows a comparison for the distribution of the shocked CO clouds and the HI gas. Here we use the HI4PI HI data instead of the GALFA-HI
data to make a plot, due to large-area bad pixels at the position of the Crab Nebula in the GALFA-HI image (see Fig.\,5).  As seen in Fig.\,6, the 
shocked CO clouds are founded at the edge of the HI shell toward the southeast of the Crab Nebula. 

It is still uncertain for the relationship between the Crab Nebula and this newly-discovered HI bubble. After checking large-field H$\alpha$ images (e.g., 
Finkbeiner 2003), we did not find extended ionized emission within the bubble. Therefore, the bubble could be opened by stellar wind from massive star(s).
Using the method suggested by Weaver et al. (1977), the value of the mechanical luminosity of the wind ($L_{\rm wind}$) can be calculated by 
\begin{equation}\label{f.luminosity}
 L_{\rm wind}~\approx ~\frac{1}{3}\left (\frac{{\it n_{\rm gas}}}{\rm cm^{-3}}\right )~\left (\frac{{\it R_{\rm bubble}}}{\rm pc}\right)^{2}~\left(\frac{{\it V_{\rm exp}}}{\rm km\,s^{-1}}\right)^{3} \times\,10^{30}~{\rm erg\,s^{-1}},
\end{equation}
in order to excavate a bubble with a radius of $R_{\rm bubble}$ and an expansion velocity of $V_{\rm exp}$ within a cloud with a density of $n_{\rm gas}$.
With the radius ($\sim$\,45\,pc) and expansion velocity ($\sim$\,5\,\kms) measured above, the $L_{\rm wind}$ is then estimated to be 
$\sim$\,8.4\,$\times$\,$\left (\frac{{\it n_{\rm gas}}}{\rm cm^{-3}}\right)$\,$\times$\,10$^{34}$\,erg\,s$^{-1}$.
The kinetic timescale $t_{\rm kin}$ of the wind needed for opening the bubble is estimated by
\begin{equation}\label{f.energy}
 t_{\rm kin}~(\rm Myr) = \frac{16}{27}\frac{\it R_{\rm bubble}}{\it V_{\rm exp}},
\end{equation}
which is roughly 5.3\,Myr. For the Crab Pulsar, the mass of the progenitor is suggested to be $\sim$\,8-13\,\Msun\ (see Hester 2008 and references therein). 
This kind of massive star is able to driving the stellar wind estimated above, given that the density $n_{\rm gas}$ surrounding the star is less than 1\,cm$^{-3}$. 
However, for the large shell with a diameter of $\sim$\,5$^\circ$, it cannot be opened by the stellar wind from the progenitor of the Crab Pulsar. Other massive
stars or supernovae are needed to supply the energy.

%%%%%%%%%%%%%%%%%%%%%%%%%%%%%%%%%%%
\section{Summary}
%%%%%%%%%%%%%%%%%%%%%%%%%%%%%%%%%%%

Using the PMO 13.7\,m telescope, we present large-field and high-sensitivity CO\,(1--0) line observations toward the Crab Nebula, 
in order to better understand the interstellar gas environment of this well-known historical remnant. The main results of this work 
are summarized below.

(1) The CO observations show a large-scale molecular cloud, extending from --2 to 6\,\kms, toward the north of the Crab Nebula. After 
measuring the distance of the large-scale cloud ($\sim$\,1374\,pc), we conclude that the cloud is located in front of the Crab Nebula. 

(2) In the deep CO observations, small-scale molecular clouds are detected in the velocity range of [0, 16]\,\kms\ toward the Crab Nebula. 
After checking the CO spectra, we find that molecular clouds in the velocity range of [5, 11]\,\kms\ show shocked signatures. These clouds, 
displaying as an arc, are located roughly 0\fdg4-0\fdg5 to the southeast of the Crab Nebula.

(3) Based on the complementary HI data, we find around the Crab Nebula a large HI bubble, with a dimension of roughly 2\fdg3\,$\times$\,2\fdg6 
and an expansion velocity of $\sim$\,5\,\kms. The shocked CO clouds are spatially distributed at the edge of the HI shell toward the southeast of 
the Crab Nebula.

(4) The combined observational data suggest that the molecular clouds at the velocity of [5, 11]\,\kms\ are affected by front shocks from the 
SN\,1054 explosion, while the HI bubble could be opened by the stellar wind from its progenitor. In this picture, the kinetic energy released
from SN\,1054 is estimated to be roughly 3.5\,$\times$\,10$^{51}$ erg. 

Nevertheless, it must be noted that it is still difficult to make a solid association between the shocked molecular clouds and the Crab Nebula, 
due to large uncertainty in the distance measurement towards molecular clouds.

%%%%%%%%%%%%%%%%%%%%%%%%%%%%%%%%%%%
%%%%%%%%%%%%%%%%%%%%%%%%%%%%%%%%%%%

\begin{acknowledgements}
We thank the referee for providing insightful suggestions, which help us to improve this work.
This research made use of the data from the Milky Way Imaging Scroll Painting (MWISP) project, which is a multiline survey in 
$^{12}$CO/$^{13}$CO/C$^{18}$O along the northern Galactic plane with the PMO 13.7\,m telescope. We are grateful to all the 
members of the MWISP working group, particularly the staff members at the PMO 13.7 m telescope, for their long-term support. 
MWISP is sponsored by the National Key R\&D Program of China with grants 2023YFA1608000, 2017YFA0402701, and the
CAS Key Research Program of Frontier Sciences with grant QYZDJ-SSW-SLH047. This work is supported by the National Natural 
Science Foundation of China (grant No. 12041305). X.C. acknowledges the support from the Tianchi Talent Program of Xinjiang 
Uygur Autonomous Region.
\end{acknowledgements}

\clearpage

% Please note that we have included the references below in
% order to compile the document, but we ask you to:

\clearpage
%%%%%%%%%%%%%% Figures %%%%%%%%%%%%%%%%%

\begin{figure*}[!htbp]
\centering
\includegraphics[width = 0.90\textwidth]{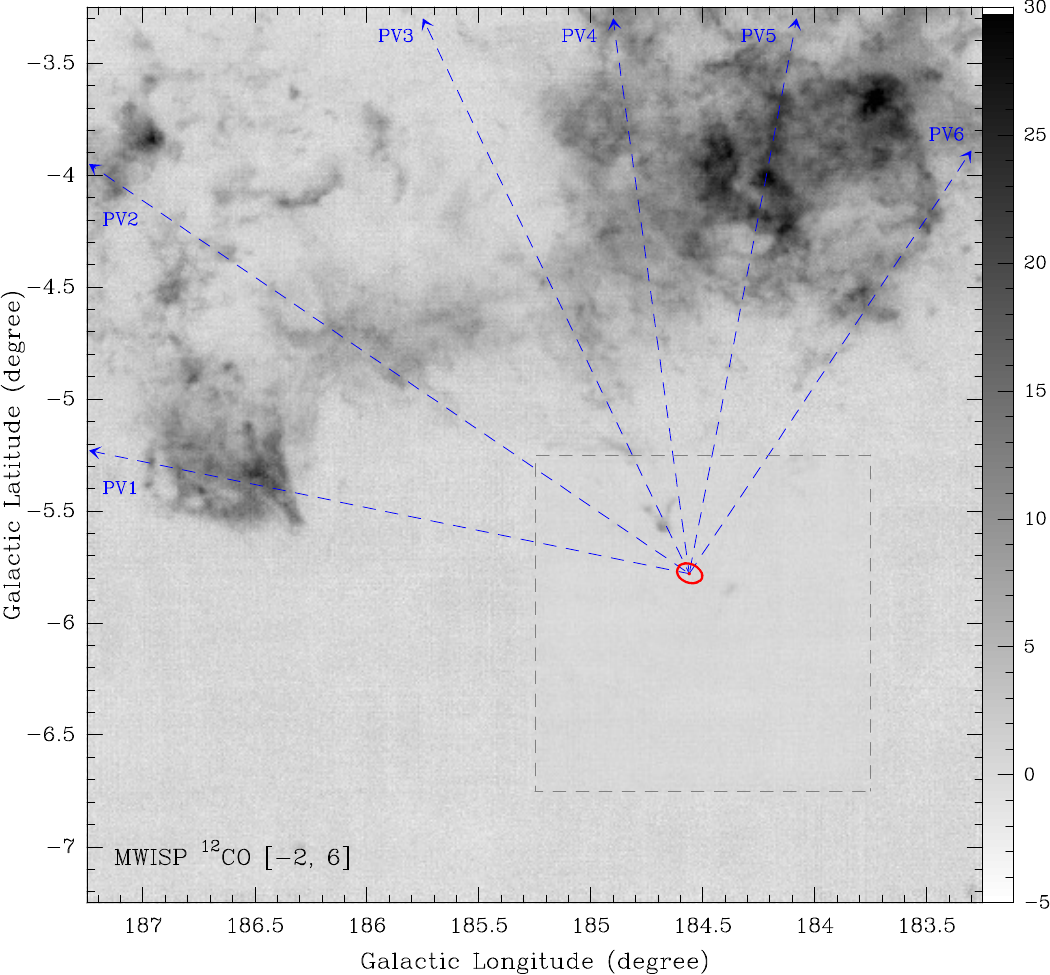}
\caption{The intensity image of the MWISP $^{12}$CO\,(1--0) emission, integrated in the velocity range from --2 to 6\,\kms. The grey dashed square 
shows the area of deep mapping observations. The red ellipse shows the position and size (7$'$\,$\times$\,5$'$) of the Crab Nebula. The blue dashed 
arrow lines show the routings of the PV diagrams in Figure~2.}
\label{MWISP_intensity_v1}
\end{figure*}

\begin{figure*}[!htbp]
\centering
\includegraphics[width = 0.9\textwidth]{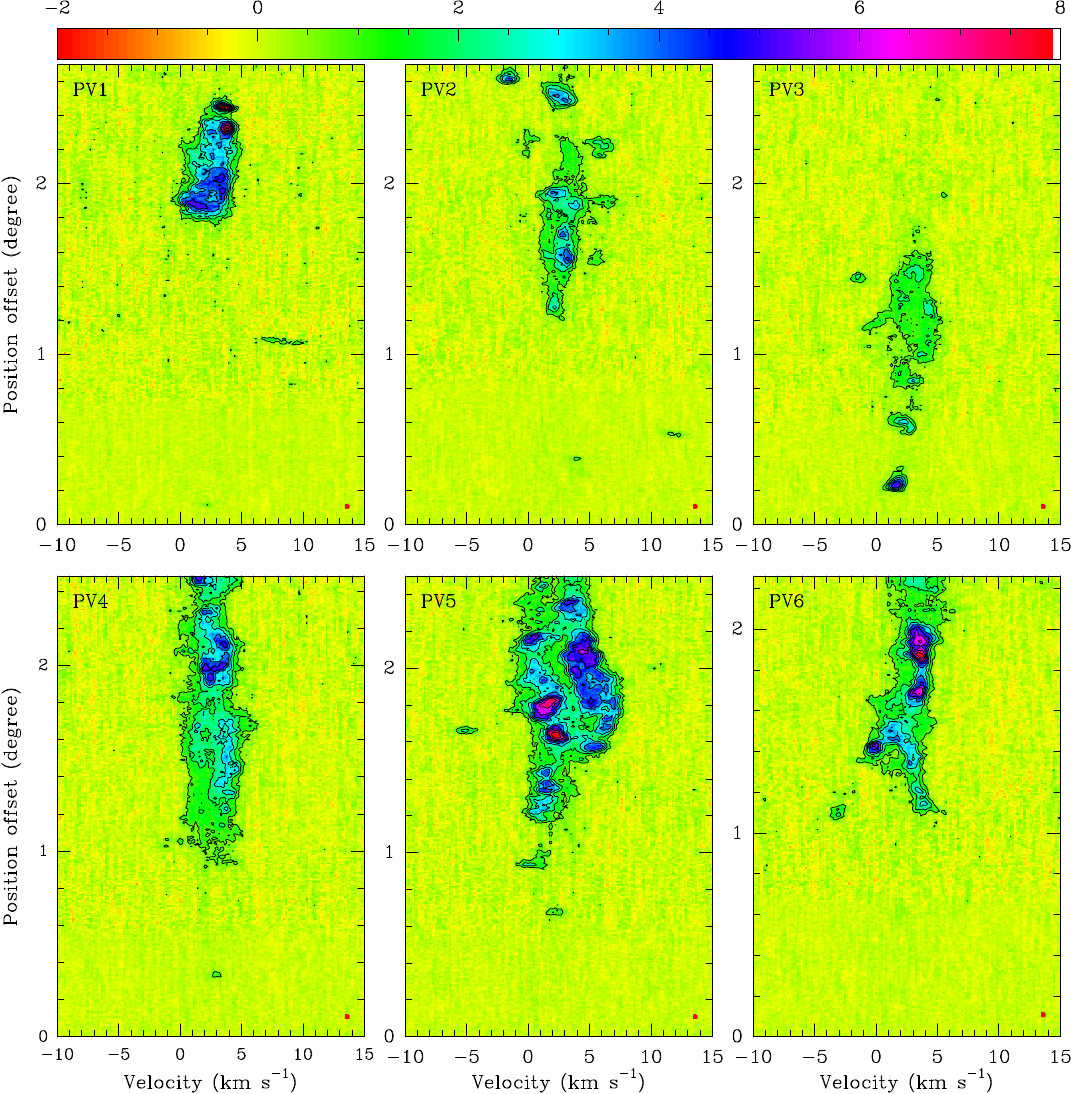}
\caption{The PV diagrams from Crab across the north molecular cloud (see routings shown in Figure~1). In each panel, the contours
start from 3\,$\sigma$ and then increase by steps of 4\,$\sigma$, where 1\,$\sigma$ in the large-field $^{12}$CO observations is roughly 0.25\,K. 
The red rectangle in the bottom right corner of each panel shows the angular (52$''$) and velocity (0.16\,\kms) resolution in the MWISP $^{12}$CO 
observations.}
\label{MWISP_PV}
\end{figure*}

\begin{figure*}[h!]
\centering
\includegraphics[width = 0.6\textwidth]{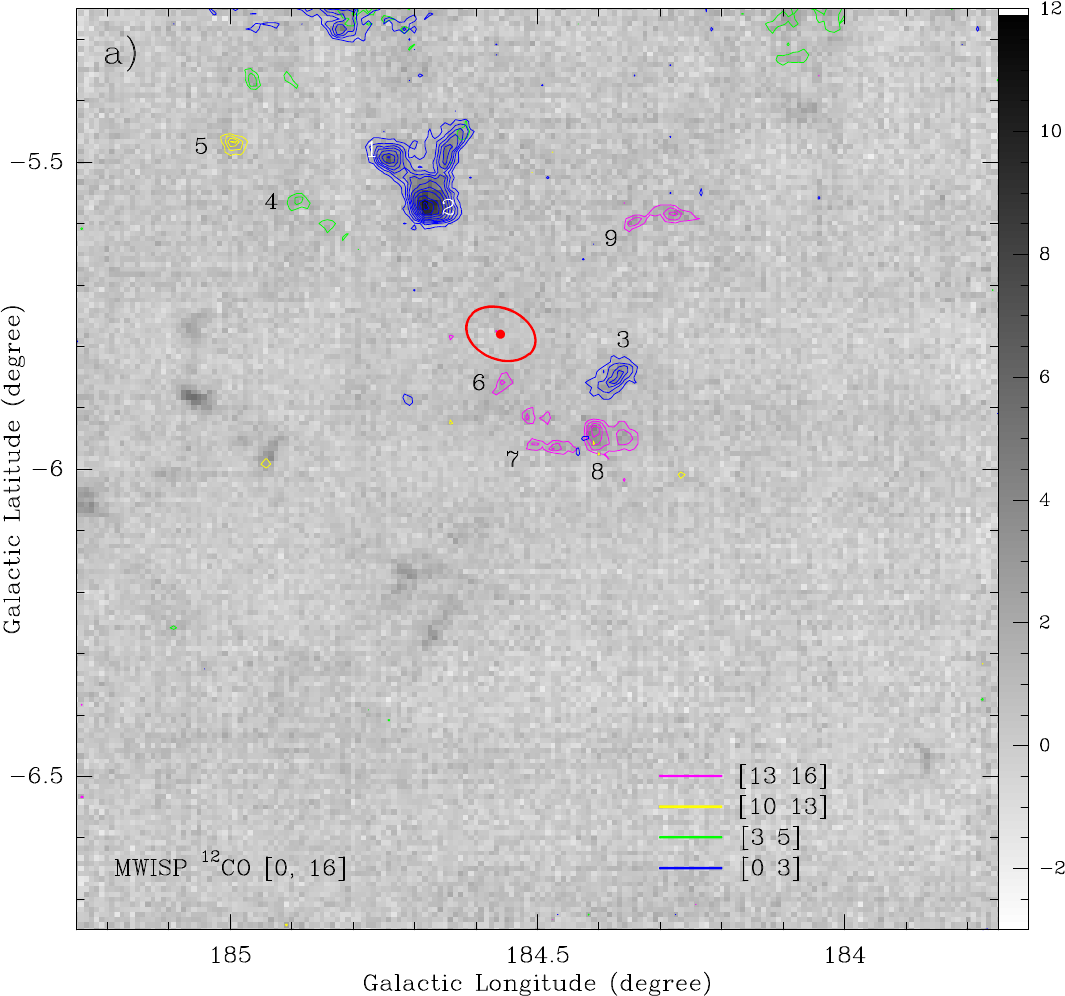}
\includegraphics[width = 0.6\textwidth]{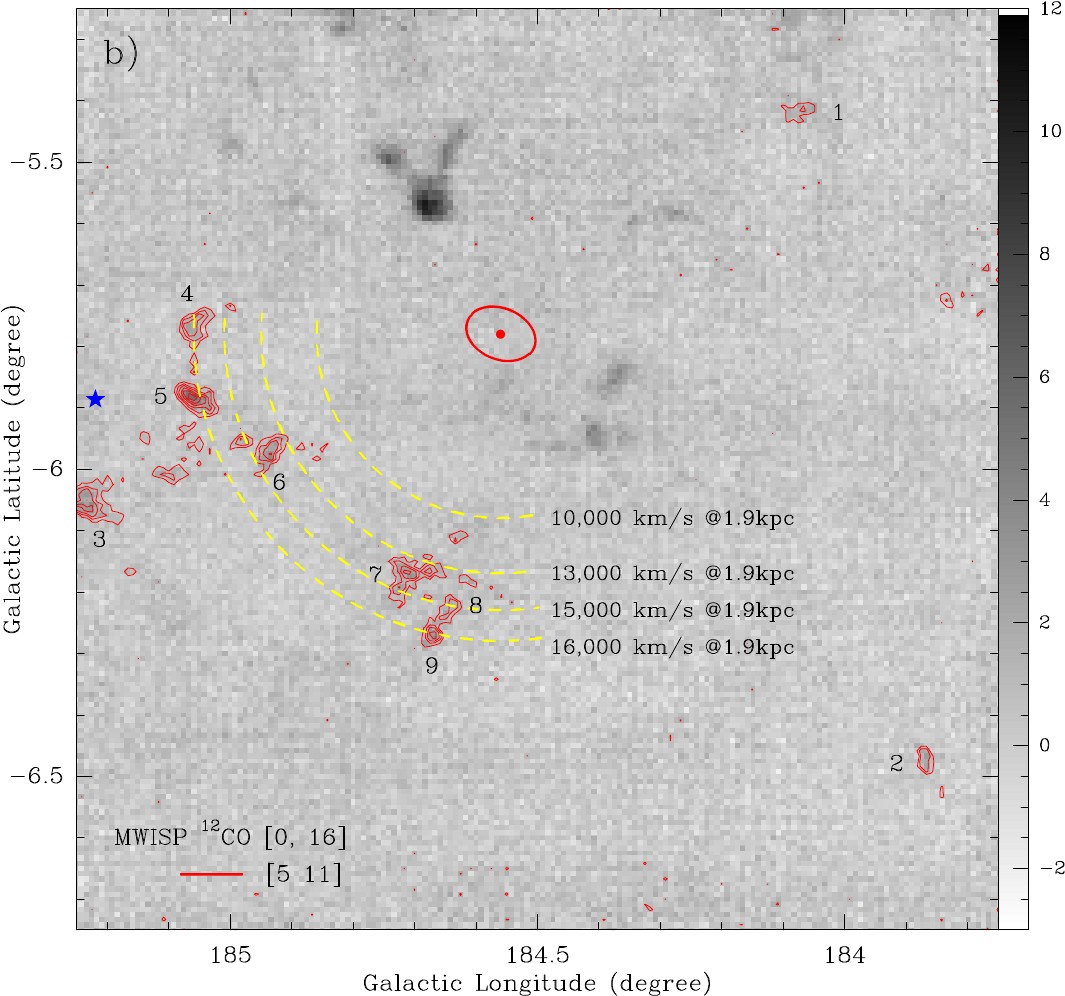}
\caption{The $^{12}$CO intensity image (grey scale) in the deep observations, integrated within the velocity range of [0, 16]\,\kms. The red ellipse shows 
the position and size (7$'$\,$\times$\,5$'$) of the Crab Nebula, while the red dot at the center of the ellipse shows the FWHM beam (52$''$) in the MWISP 
$^{12}$CO observations at the position of the pulsar ($l$\,=\,184\fdg56, $b$\,=\,--5\fdg78). 
(a) The overlapped different color contours show the CO emission for the clouds integrated within individual velocity ranges, starting from 4\,$\sigma$, 7\,$\sigma$, 
10\,$\sigma$, 13\,$\sigma$, 17\,$\sigma$, and then increasing with steps of 5$\sigma$, where 1\,$\sigma$ corresponds to $\sim$\,0.23\,K\,\kms\ for the [0, 3]\,\kms\ 
(blue) and [10, 13] (yellow), $\sim$\,0.19\,K\,\kms\ for the [3, 5] (green) and [13, 16] (pink) velocity ranges, respectively. The numbers mark the positions of the clouds 
to sample CO spectra (see Figure~A.4). 
(b) The overlapped contours show the CO emission for the clouds integrated within the velocity range of [5, 11]\,\kms, starting from 3\,$\sigma$, 5\,$\sigma$, 
7\,$\sigma$, and then increasing with steps of 3$\sigma$, where 1\,$\sigma$ corresponds to $\sim$\,0.33\,K\,\kms. The numbers mark the positions of the 
clouds to sample CO spectra (see Figure~4). The blue star marks the position of the O7 star SAO\,77293 (also known as HD\,36879). Yellow dashed arcs show 
the putative outer shocks from SN\,1054 at different velocities.}
\label{MWISP_deep1}
\end{figure*}

\begin{figure*}[!htbp]
\centering
\includegraphics[width = 0.9\textwidth]{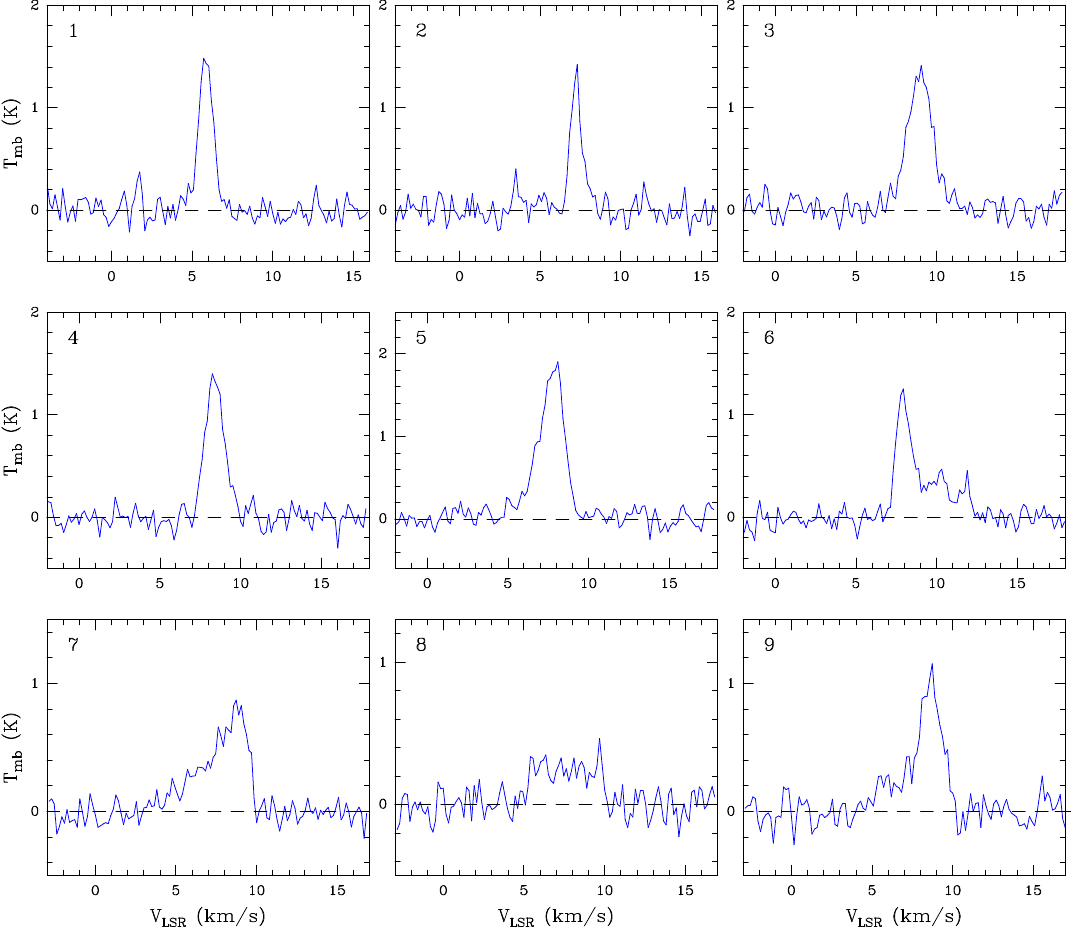}
\caption{The $^{12}$CO spectra of the clouds in the velocity range of [5, 11]\,\kms. All the spectra are sampled in a circle with a radius of 90$''$, 
and the sampled positions are marked in Figure 3b.}
\label{MWISP_spectra_v2}
\end{figure*}

\begin{figure*}[!htbp]
\centering
\includegraphics[width = 0.80\textwidth]{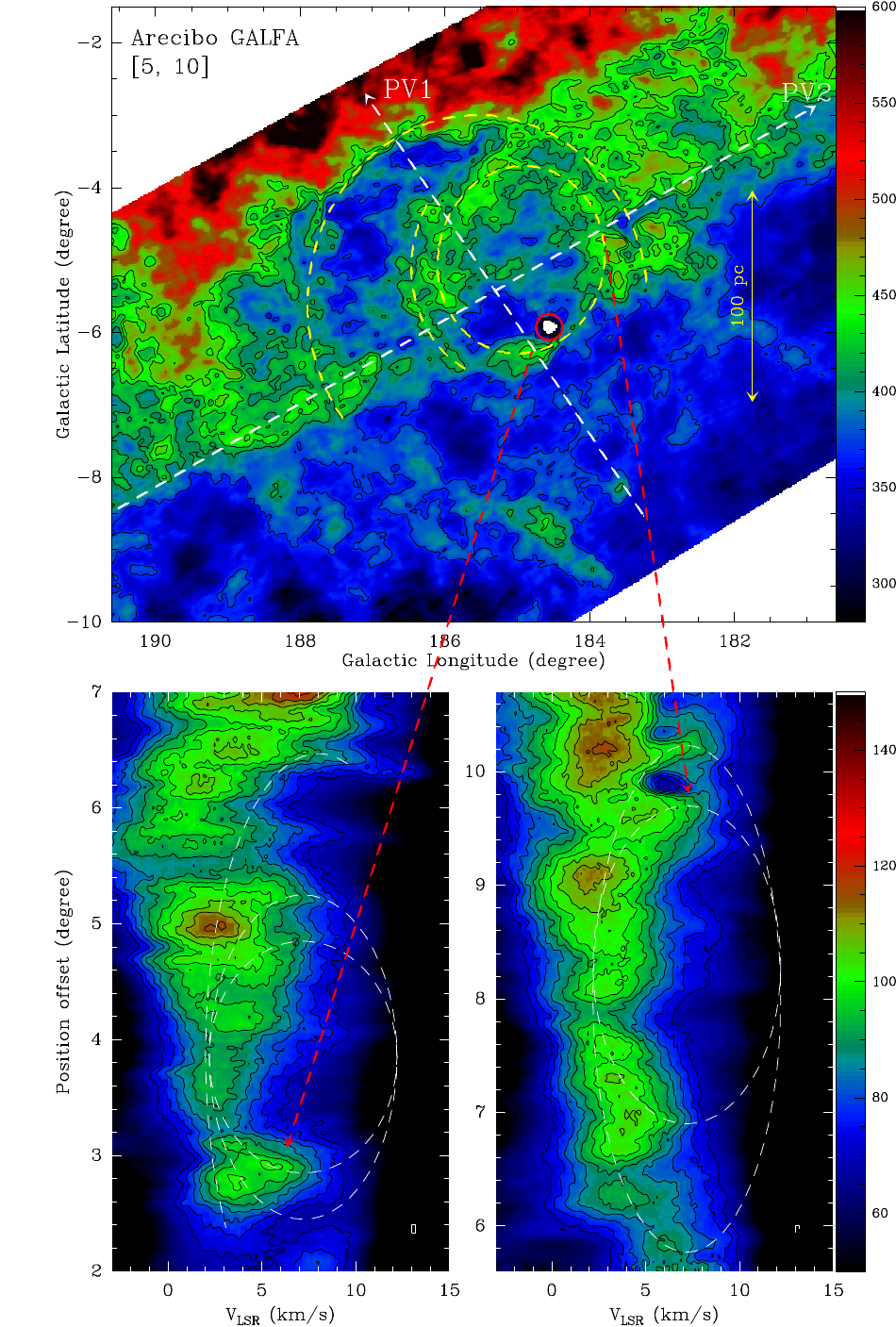}
\caption{The top panel shows the GALFA-HI intensity image, integrated in the velocity range of [5, 10]\,\kms. The contours correspond to 370, 410, 430, and 450\,K\,\kms. 
The red circle shows position of the Crab Nebula. The yellow dashed circles show the bubble and shells distinguished in the HI image, while white dashed arrows show 
the routings of the PV diagrams shown in the bottom panels. For the two PV diagrams, the contours start from 70\,K and then increase by steps of 5\,K. The white (partial) 
ellipses show the fittings toward the cavity-like (spur-like) structures. The white rectangle in the bottom right corner of the panels shows the angular (4$'$) and velocity 
(0.184\,\kms) resolutions in the GALFA-HI survey.}
 \label{GALFA_intensity_v1}
\end{figure*}

\begin{figure*}[!htbp]
\centering
\includegraphics[width = 0.7\textwidth]{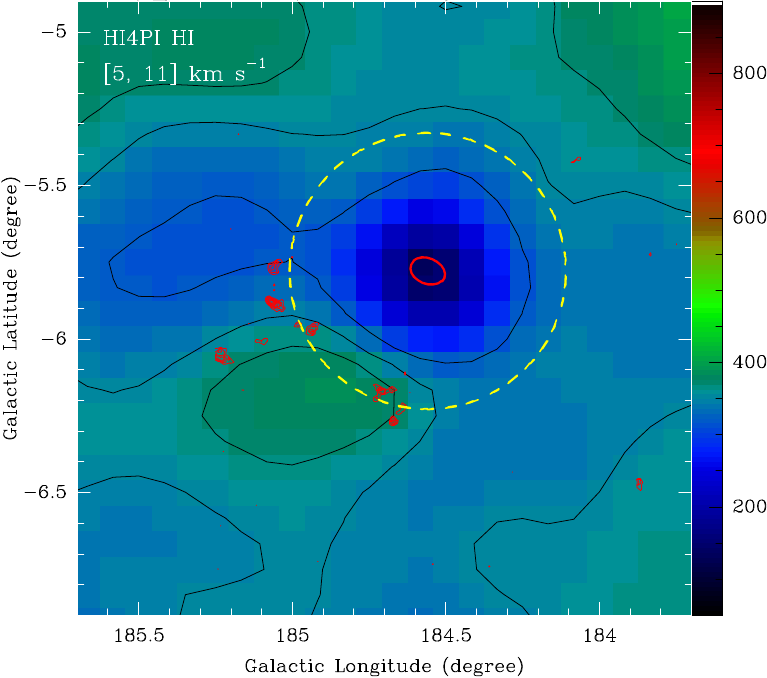}
\caption{The HI4PI HI intensity image, integrated in the velocity range of [5, 11]\,\kms. The contours start at 320\,K\,\kms\ and then increase by steps of 20\,\kms. The red 
contours show the CO emission in the velocity range of [5, 11]\,\kms\ (from Fig\,3b), while the yellow dashed circle (radius of 0\fdg45) show the putative outer shocks from 
SN 1054 at the velocity of 15,000\,\kms.}
\label{HI4PI_intensity_v1}
\end{figure*}

\clearpage

%%%%%%%%%%%%%%%%%%%%%%%%%%%%%%%%%%%%%%%%%%%%%%%%%%%%%%%%%%%%%%%
% Appendices must be placed after   \end{thebibliography}
% They will be placed automatically on a new page.
%%%%%%%%%%%%%%%%%%%%%%%%%%%%%%%%%%%%%%%%%%%%%%%%%%%%%%%%%%%%%%%

\begin{appendix}
%%%%%%%%%%%%%%%%%%%%%%%%%%%%%%%%%%%%%%%%%%%%%%%%%%%%%%%%%%%%%%%
%____________________________________________________________
%       Wide floats at the start of an appendix: first method
%-------------------------------------------------------------
% To prevent a blank page after the start of an appendix:
% - Switch to one \onecolumn first
% - Declare the section title
% - Declare the onecolumn float with the parameter [h!]
% - Revert to \twocolumn at the end of the section
\onecolumn

\section{Large-field and deep MWISP CO observations}

As we introduced in Section~2, the typical rms noise in the MWISP survey was about 0.5\,K for the $^{12}$CO line (at a velocity resolution of $\sim$\,0.16\,\kms). 
To obtain higher sensitivity, we deeply mapped a 1\fdg5\,$\times$\,1\fdg5 area (183\fdg75\,$\leq$\,$l$\,$\leq$\,185\fdg25, 
$-$6\fdg75\,$\leq$\,$b$\,$\leq$\,$-$5\fdg25) around the Crab Nebula, in which the rms noise was about 0.25\,$\pm$\,0.03\,K for $^{12}$CO. 
Figure~A.1 shows the rms distribution in the MWISP observations. Figure~A.2 shows the large-field (4$^\circ$\,$\times$\,4$^\circ$) $^{12}$CO velocity channel 
maps toward the Crab Nebula. Figure~A.3 shows the $^{12}$CO velocity channel maps toward the Crab Nebula in the deep (1\fdg5\,$\times$\,1\fdg5) observations.
Figure~A.4 shows the sampled $^{12}$CO spectra.

\begin{figure*}[!h]
\centering
\includegraphics[width = 0.7\textwidth]{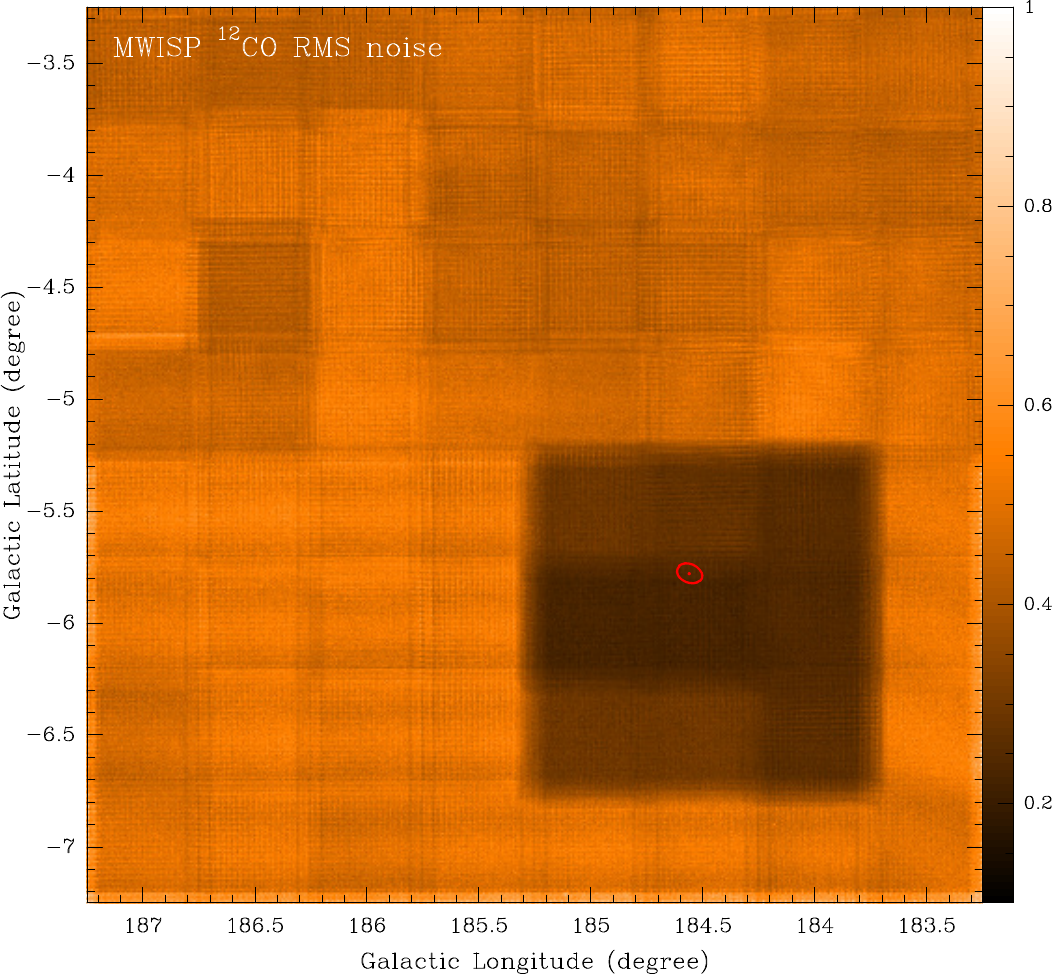}
\caption{The distribution of rms noises in the MWISP $^{12}$CO\,(1--0) observations (velocity resolution of $\sim$\,0.16\,\kms) toward the Crab Nebula. 
The unit of the scale bar is K. The relatively dark region shows the deep mapping area in the CO observations. The red ellipse shows the position and 
size (7$'$\,$\times$\,5$'$) of  the Crab Nebula, while the red dot at the center of the remnant shows the FWHM beam (52$''$) in the MWISP $^{12}$CO 
observations at the position of the pulsar ($l$\,=\,184\fdg56, $b$\,=\,--5\fdg78).}
\label{MWISP_rms}
\end{figure*}

\begin{figure*}[!htbp]
\center
\includegraphics[width = 0.95\textwidth]{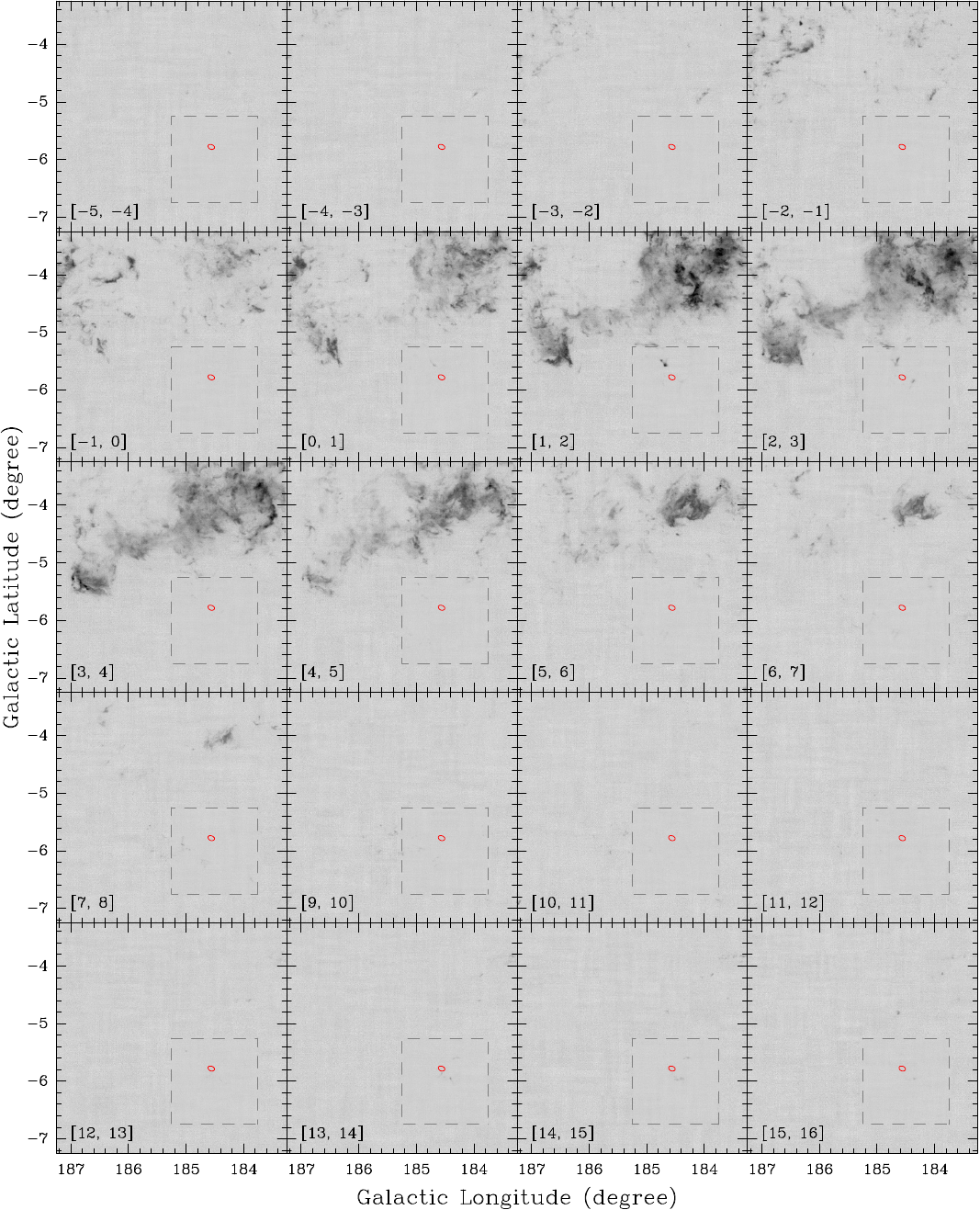}
\caption{The velocity-integrated intensity channel maps of the MWISP $^{12}$CO\,(1--0) emission in the large-field observations. The integrated velocity 
range is written in the bottom left corner of each panel (in km\,s$^{-1}$). The plotting scale for each panel is same (from --2 to 9\,K\,\kms). In each panel, 
the grey dashed square shows the area of deep mapping observations,  while the red ellipse shows the position and size (7$'$\,$\times$\,5$'$) of the 
Crab Nebula.}
\label{MWISP_channel_v1}
\end{figure*}

\begin{figure*}[!htbp]
\centering
\includegraphics[width = 0.95\textwidth]{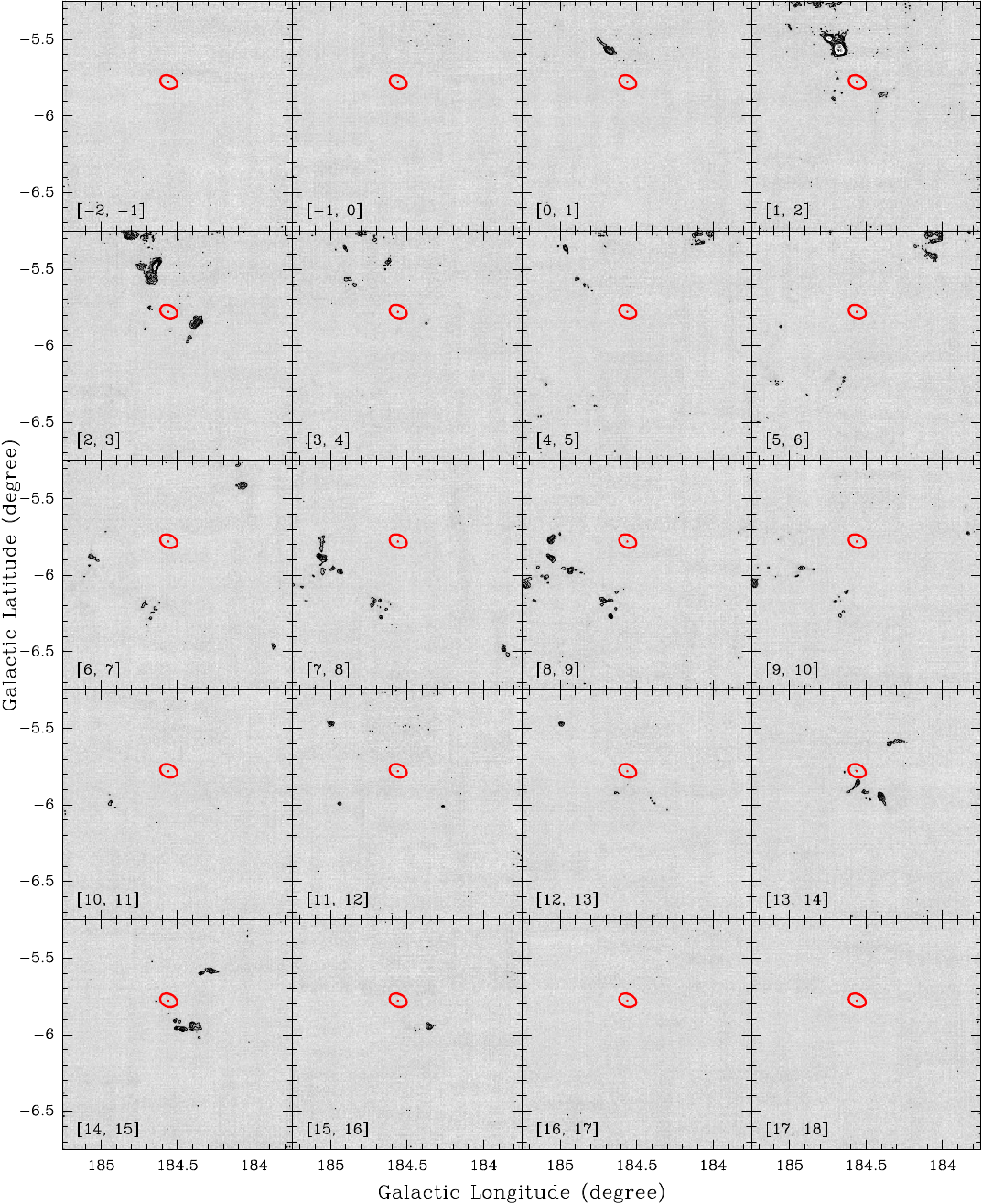}
\caption{The velocity-integrated intensity channel maps of the MWISP $^{12}$CO\,(1--0) emission in the deep observations. In each panel, contour 
levels correspond to 3, 5, 7, 10, 14\,$\sigma$, then increase in steps of 5\,$\sigma$, where the 1\,$\sigma$ level is $\sim$\,0.16\,K\,km\,s$^{-1}$. The 
integrated velocity range is written in the bottom left corner of each panel (in km\,s$^{-1}$). The red ellipse shows the position and size 
(7$'$\,$\times$\,5$'$) of the Crab Nebula, while the red dot at the center of the ellipse shows the FWHM beam (52$''$) in the MWISP $^{12}$CO 
observations at the position of the pulsar ($l$\,=\,184\fdg56, $b$\,=\,--5\fdg78).}
\label{MWISP_channel_v2}
\end{figure*}

\begin{figure*}[!htbp]
\centering
\includegraphics[width = 0.90\textwidth]{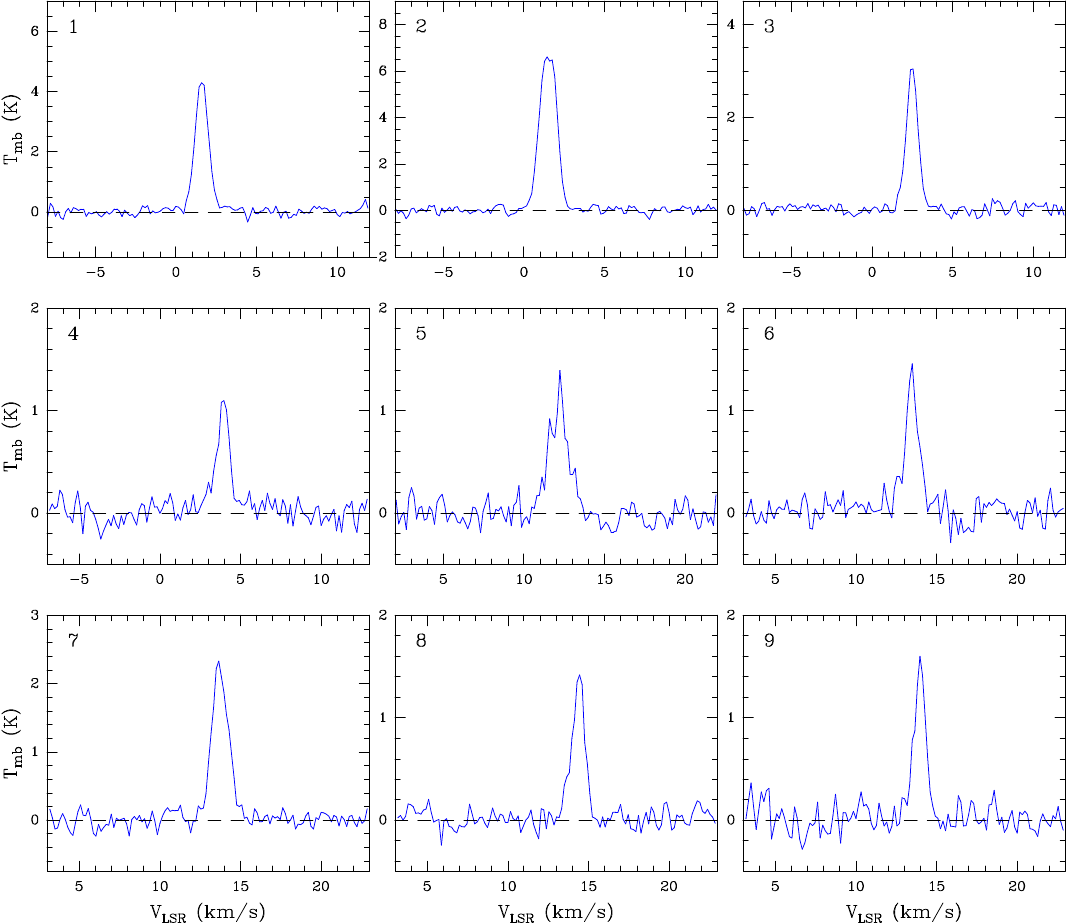}
\caption{The $^{12}$CO spectra of the clouds in the velocity ranges of [0, 3]\,\kms\ (spectra No. 1--3), [3, 5]\,\kms\ (spectrum No. 4), [11, 13]\,\kms\ 
(spectrum No. 5), and [13, 16]\,\kms\ (spectra No. 6--9). All the spectra are sampled in a circle with a radius of 90$''$, and the sampled positions 
are marked in Figure 3a.}
\label{MWISP_spectra_v1}
\end{figure*}

\section{The distances of the molecular clouds}

Based on the MWISP $^{12}$CO and $Gaia$ DR3 data (Gaia et al. 2023), we measure the distances toward the molecular clouds in the observed region, 
using the same method as described in Yan et al. (2021). 
As seen in Figure~B.1., a large-scale molecular cloud is seen to the northeast of the Crab Nebula. The distance measured for this molecular cloud is 
$\sim$\,1374\,pc. Therefore, this cloud is located in front of the Crab Nebula.
For the shocked molecular clouds to the southeast of the Crab Nebula (see Figure~B.2), larger uncertainty remains in the distance measurement, due to the 
small angular sizes and column densities of these clouds. The estimated distance of these clouds is roughly 1800\,pc.

\begin{figure*}
\centering
\includegraphics[width = 0.85\textwidth]{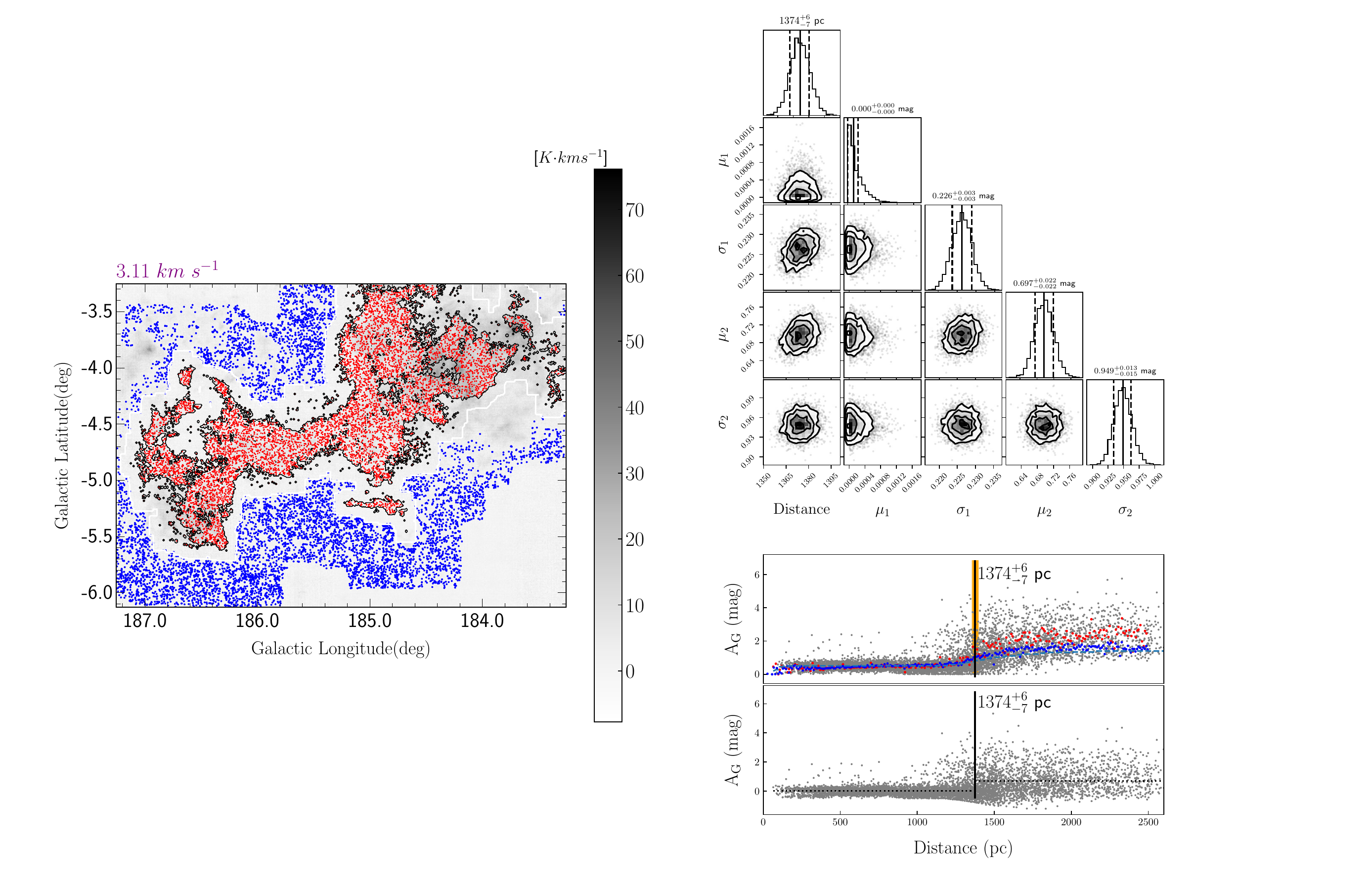}
\caption{The measured distance toward the north large-scale cloud. The left panel shows the MWISP $^{12}$CO intensity image. The black contour 
shows the edge of molecular cloud (3\,$\sigma$ threshold), while red and blue dots represent on- and off-cloud $Gaia$ DR3 stars. In the right bottom 
panel, red and blue points represent on- and off-cloud stars (binned every 5\,pc), respectively. The dashed green line is the modeled extinction $A_{G}$. 
The distance is derived with raw on-cloud $Gaia$ DR3 stars, which are represented with gray points. The black vertical lines indicate the distance 
($D$) estimated with Bayesian analyses and Markov Chain Monte Carlo (MCMC) sampling, and the shadow area depicts the 95\% highest posterior 
density (HPD) distance range. The plots of the MCMC samples are displayed in the right top panel (distance, the extinction of foreground and background 
stars and uncertainties). The mean and 95\% HPD of the samples are shown with solid and dashed vertical lines, respectively.}
\label{MWISP_distance1}
\includegraphics[width = 0.85\textwidth]{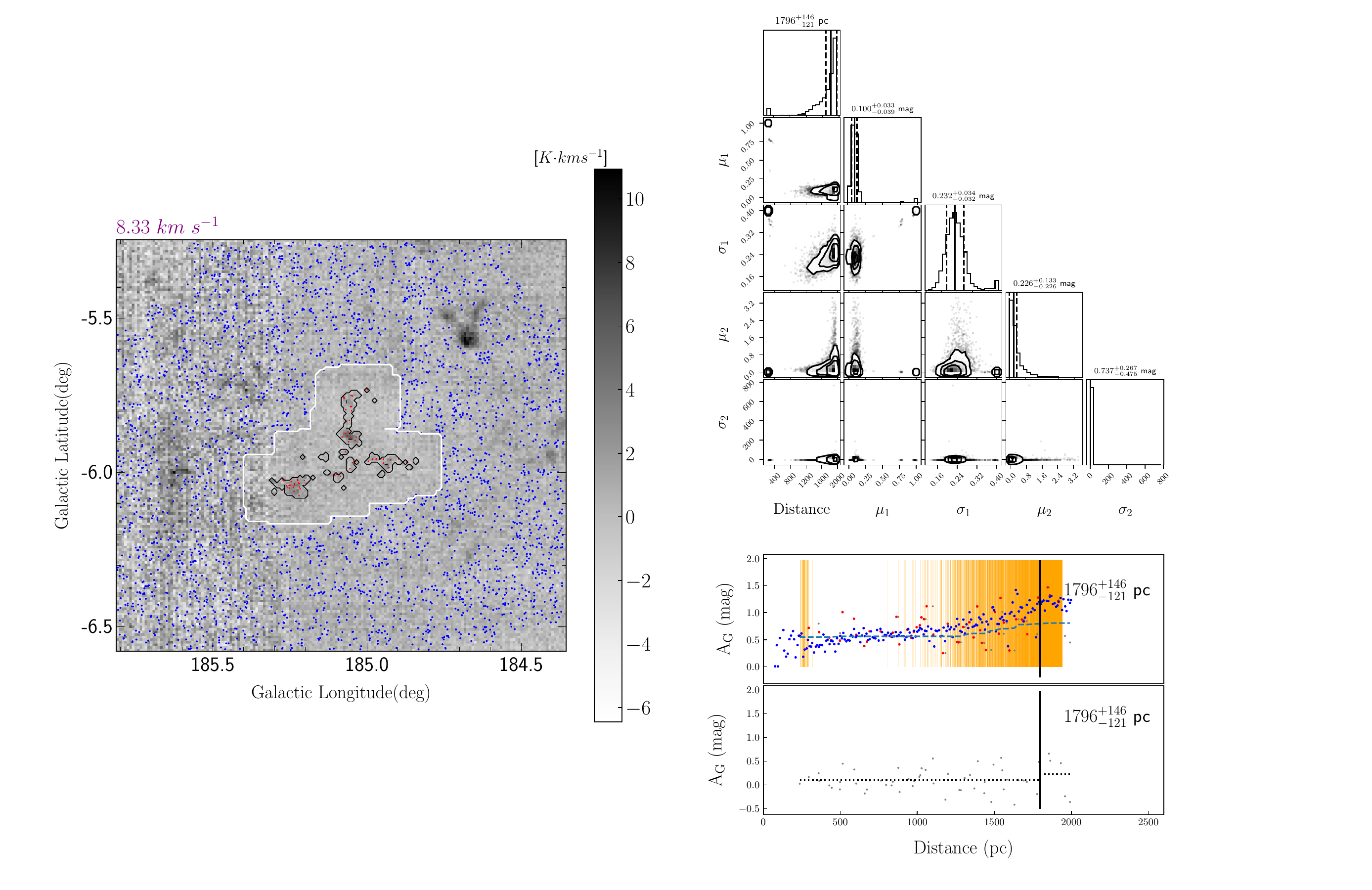}
\caption{The same as the Figure~B1, but the clouds at the velocity range of [5, 11]\,\kms.}
\label{MWISP_distance2}
\end{figure*}

\clearpage
\section{The CO spectra of the shocked molecular clouds}

Figure~C.1 shows the $^{12}$CO intensity from Figure~3b, while Figure~C.2 shows the grid $^{12}$CO spectra superposed on the intensity images. 

\begin{figure*}[!h]
\centering
\includegraphics[width = 0.7\textwidth]{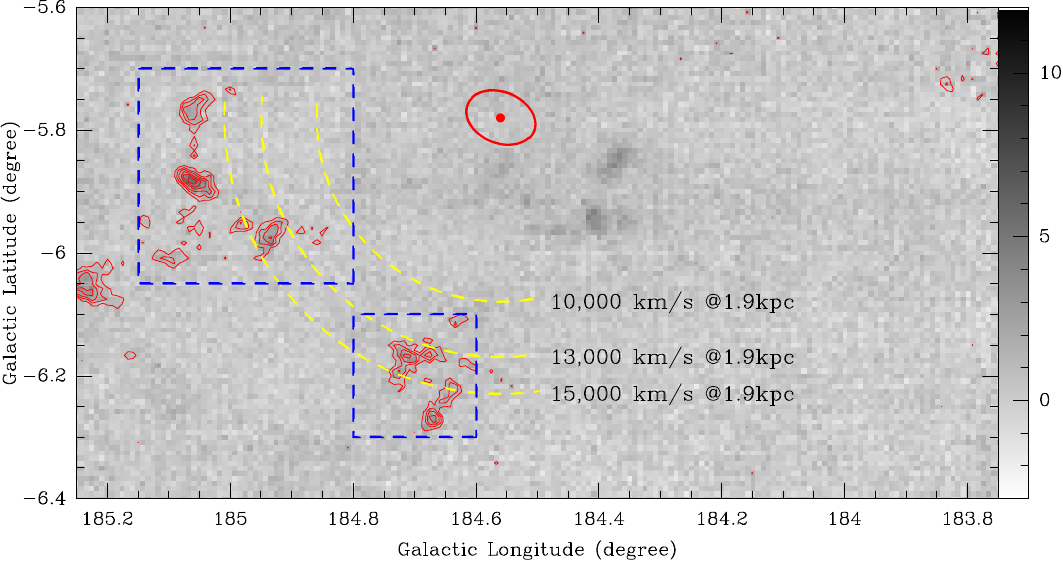}
\caption{The $^{12}$CO velocity-integrated intensity image (same as the Fig.\,3b). The two blue dashed-line rectangles mark the regions for showing the $^{12}$CO 
grid spectra.}
\label{MWISP_intensity_appendix}
\end{figure*}

\begin{figure*}[!htbp]
\centering
\includegraphics[width = 0.70\textwidth]{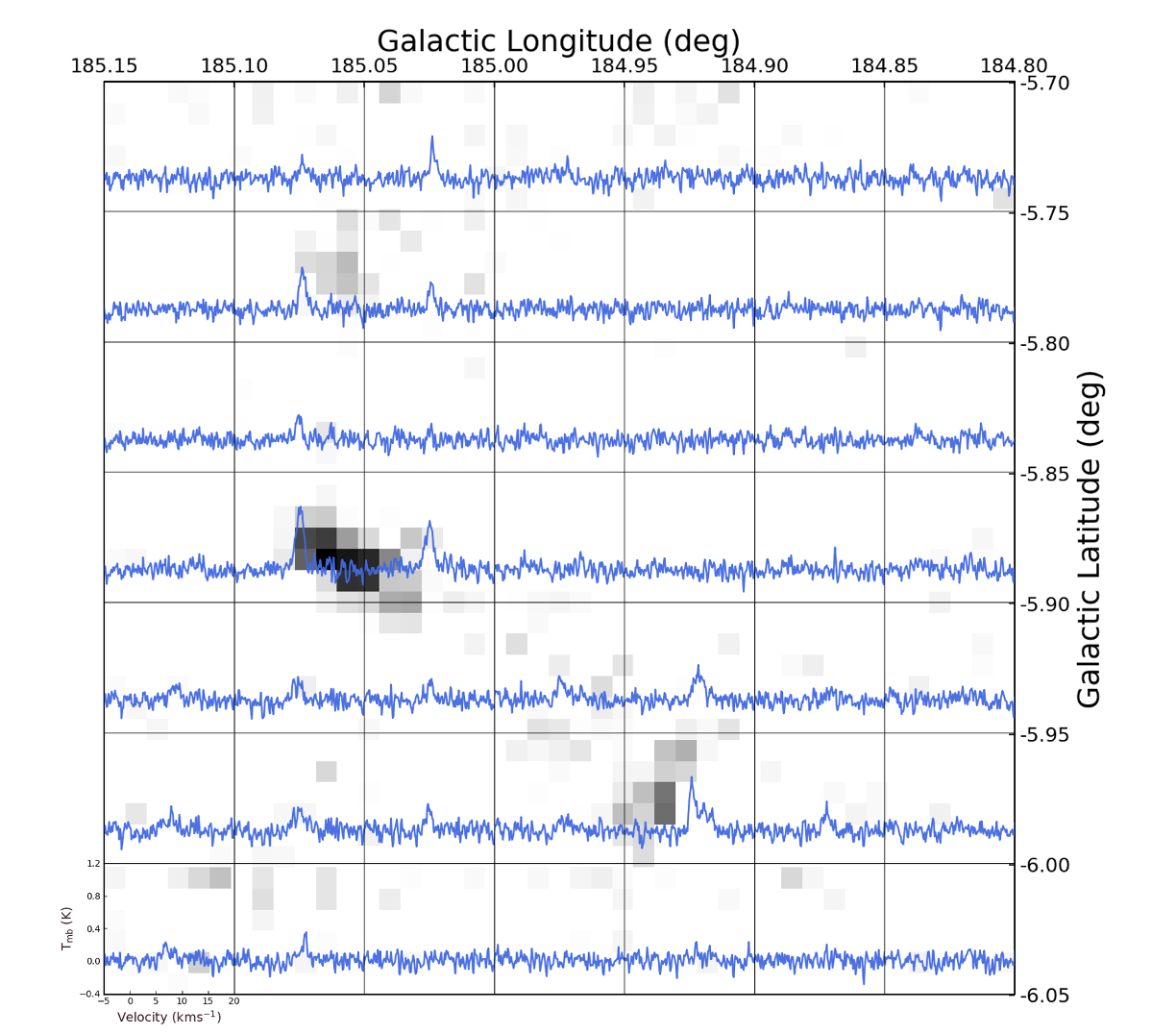}
\includegraphics[width = 0.68\textwidth]{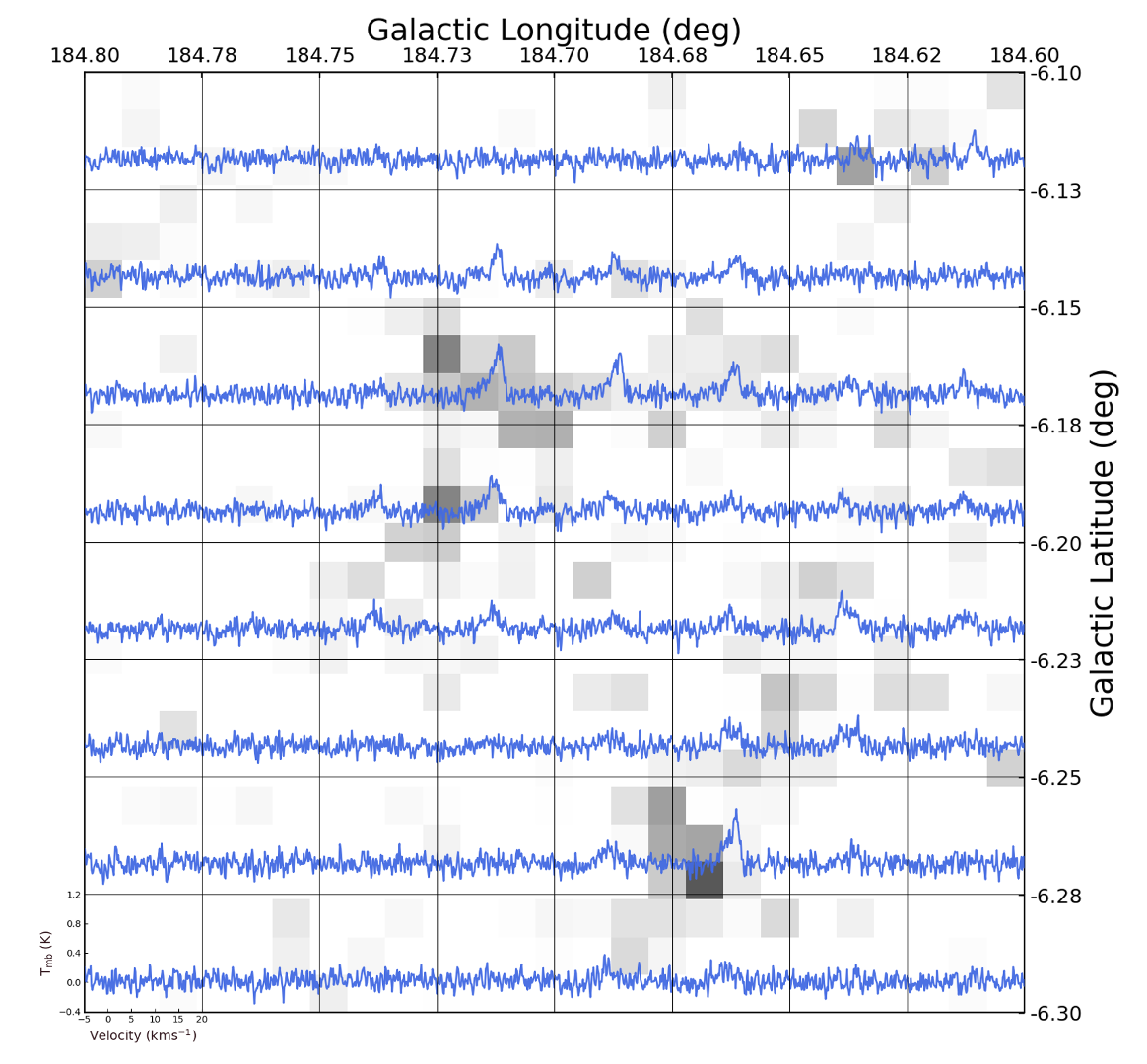}
\caption{The grid $^{12}$CO spectra of the regions marked in Fig.\,C.1, superposed on the $^{12}$CO intensity images. The temperature and velocity ranges are shown 
in the bottom-left panel for the two images. The grid spacing is 3$'$ for the top image and 1.5$'$ for the bottom image.}
\label{MWISP_appendix_spectral}
\end{figure*}

\clearpage
\section{The GALFA-HI channel maps and intensity image}

Figure~D.1 shows the GALFA-HI channel maps in the velocity range from 0 to 12\,\kms, in which detailed information about the distribution
and kinematics of the HI gas can be found. Figure~D.2 shows the GALFA-HI intensity image integrated within the [-20, 0]\,\kms. 

\begin{figure*}[!htbp]
\centering
\includegraphics[width = 0.90\textwidth]{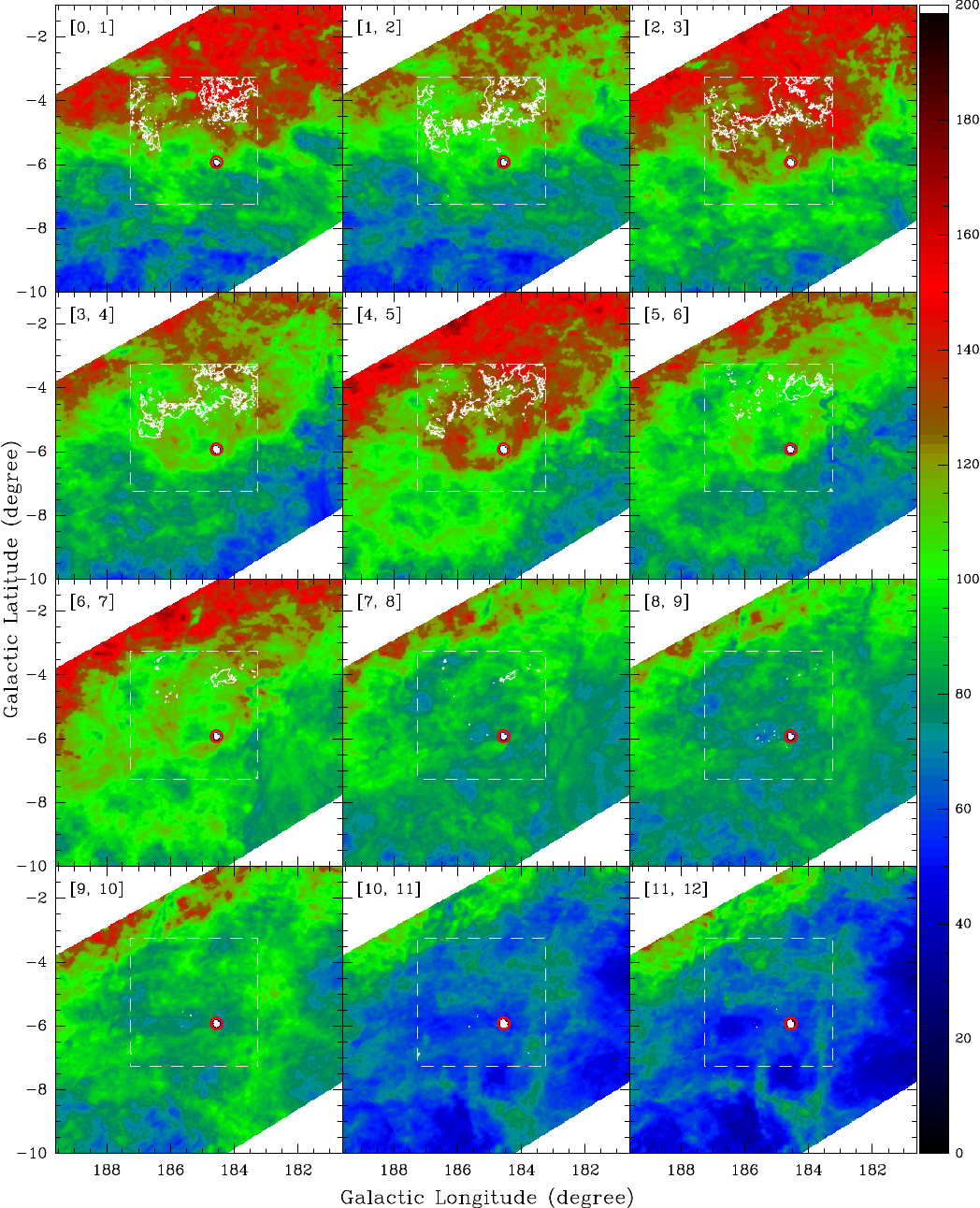}
\caption{The velocity-integrated intensity channel maps of the GALFA-HI emission. The integrated velocity range is written in the left top corner of each panel 
(in km\,s$^{-1}$). The unit of the scale bar is K\,\kms. For all panels, red circle marks the position of the Crab Nebula. The overlapped white contour (1.5\,K\,\kms) 
shows the MWISP $^{12}$CO intensity image integrated within the same velocity range as the HI emission.}
\label{GALFA_appendix_channel2}
\end{figure*}

\begin{figure}[!h]
\centering
\includegraphics[width = 0.7\textwidth]{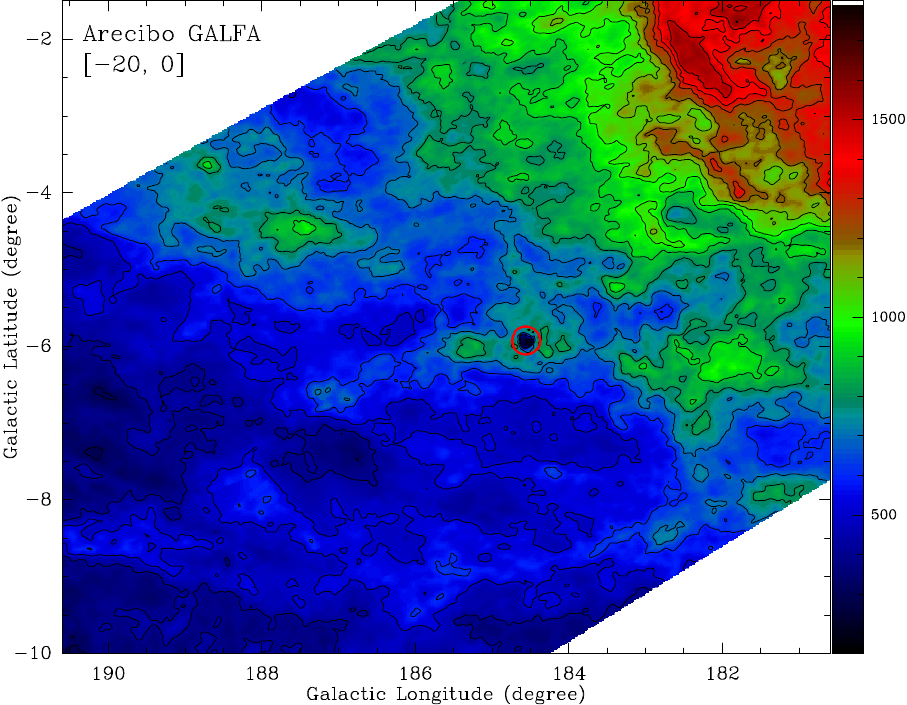}
\caption{The GALFA-HI intensity image integrated in the velocity range of [-20, 0]\,\kms. The unit of the scale bar is K\,\kms. The contours start from 
200\,K\,\kms, and then increase by steps of 100\,K\,\kms. The red circle shows the position of the Crab Nebula.}
\label{GALFA_appendix1}
\end{figure}

\end{appendix}

\end{document}